\documentclass[prb,twocolumn,superscriptaddress]{revtex4}
\usepackage{epsfig}
\usepackage{amsmath}
\usepackage{amsfonts}
\usepackage{amssymb}

\usepackage[usenames]{color}

\usepackage{ifpdf}
\ifpdf
\usepackage{epstopdf}
\fi

\newcommand{\abs}[1]{\ensuremath{\left\vert#1\right\vert}}
\newcommand{\ket}[1]{\ensuremath{\vert#1\rangle}}
\newcommand{\bra}[1]{\ensuremath{\langle #1\vert}}

\newcommand{\kb}[2]{\ensuremath{\vert #1 \rangle \langle #2 \vert}}

\renewcommand{\sp}[0]{\ensuremath{\mathbf{\sigma}_{+}}}
\newcommand{\sm}[0]{\ensuremath{\mathbf{\sigma}_{-}}}

\newcommand{\an}[1]{\ensuremath{#1}}
\newcommand{\cre}[1]{\ensuremath{#1^\dagger}}

\newcommand{\ddt}[0]{\frac{\mathrm{d}}{\mathrm{d}t}}

\renewcommand{\vec}[1]{\ensuremath{\mathbf{#1}}}

\begin{document}

\title{High fidelity all-optical control of quantum dot spins: \\ detailed study of the adiabatic approach}
\author{Erik M Gauger}
\email{erik.gauger@materials.ox.ac.uk}
\affiliation{Department of Materials, University of Oxford, OX1 3PH, United Kingdom}
\author{Simon C Benjamin}
\affiliation{Department of Materials, University of Oxford, OX1 3PH, United Kingdom}
\author{Ahsan Nazir}
\affiliation{Centre for Quantum Dynamics, Griffith University, Brisbane, Queensland 4111 Australia }
\affiliation{Centre for Quantum Computer Technology, Australia }
\author{Brendon W Lovett}
\affiliation{Department of Materials, University of Oxford, OX1 3PH, United Kingdom}


\begin{abstract}
Confined electron spins are preferred candidates for embodying quantum information in the solid state. A popular idea is the use of optical excitation to achieve the ``best of both worlds'', i.e. marrying the long spin decoherence times with rapid gating. Here we study an all-optical adiabatic approach to generating single qubit phase gates. We find that such a gate can be extremely robust against the combined effect of all principal sources of decoherence, with an achievable fidelity of $0.999$ even at finite temperature. Crucially this performance can be obtained with only a small time cost: the adiabatic gate duration is within about an order of magnitude of  a simple dynamic implementation. An experimental verification of these predictions is immediately feasible with only modest resources.
\end{abstract}

\maketitle


\section{Introduction}

Electron spins in quantum dots (QDs) are promising candidates for quantum computation (QC) due to their long intrinsic decoherence times.~\cite{loss98} Optical control of such spins, via auxiliary exciton (electron-hole) states, offers the promise of fast gating times. The `figure of merit', i.e. the ratio of decoherence time to gate time, would then be extremely high. This exciting possibility has been discussed in several proposals recently.~\cite{nazir04, calarco03,  lovett05, chen04, roszak05, lovett06}

However, direct exploitation of the excitonic degree of freedom, which we term {\it dynamic optical control}, may adversely affect the spin coherence: During the gate operation the quantum information is partially carried by the excitons, which are subject to aggressive decoherence. It has been suggested that {\it adiabatic control} could avoid this problem, by ensuring that the qubit remains encoded in low-lying states throughout the process.~\cite{calarco03, lovett05} In this paper we present calculations for both forms of gate under the {\it combined} effect of the principal decoherence mechanisms, photon emission from excitonic recombination and acoustic phonon interaction. In contrast to previous work,~\cite{calarco03, lovett05} we derive a full master equation (ME) solution for the gate dynamics that simultaneously incorporates all decoherence channels.

Our principle interest is in evaluating the performance of the adiabatic approach, and the Markovian master equation technique that we develop is tailored for this purpose. We will show that, although the adiabaticity condition introduces the possibility of Landau-Zener transitions, this does not significantly impair gate performance. Error rates less than $10^{-3}$ can be achieved even in the presence of multiple decoherence channels. In order to provide a meaningful context for these results, we will also consider the performance of a simple dynamic approach in the same parameter regime i.e. relatively weak driving. However this form of gate is not the primary focus of our study, and we will neglect more advanced dynamic models and techniques such as pulse shaping,~\cite{hohenester04} and also the effects of pure dephasing that may become significant in the case of ultrafast addressing.~\cite{alicki04}

Our predictions for both the dynamic and the adiabatic gates can be tested immediately; experiments would only involve a single QD and one laser.~\cite{ramsay07}
Such a demonstration would provide an important step towards implementing more general gates based on adiabatic approaches.~\cite{chen04, caillet07, grodecka07}

\section{Model}

\begin{figure}
\begin{center}
\includegraphics[width=0.45\textwidth]{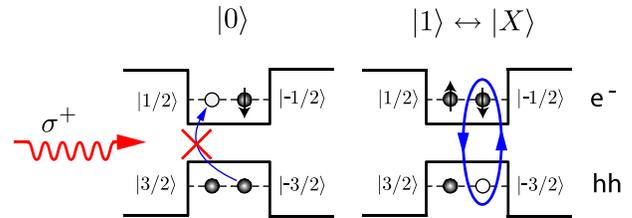}
\caption{(Color online) Pauli blocking effect: heavy holes occupy the lowest energy valence band states with spin $m_z = \pm 3/2$ whereas light holes have spin $m_z = \pm 1/2$ at a slightly higher energy. Therefore, $\sigma^+$ light cannot excite electron hole pairs if the qubit is in the state \ket{0} with $m_z = -1/2$. On the other hand, if the qubit is in the \ket{1} configuration the excitation is possible leading to a three particle trion \ket{X} state. }
\label{fig:pauli_blocking}
\end{center}
\end{figure}

Consider a  self-assembled QD that is doped such that one excess electron, the spin of which is the qubit, permanently occupies the lowest energy state of the conduction band. If the dot is irradiated with $\sigma^{+}$ polarized laser light only one of the electron spin configurations is compatible with the creation of an additional exciton due to the Pauli blocking effect \cite{calarco03, nazir04} (illustrated in Fig \ref{fig:pauli_blocking}). The resulting three particle state is called a trion and denoted \ket{X}. In this case, the logical \ket{0} state (defined as spin down \ket{\downarrow}) is unaffected by the laser pulse. Exploiting the selective coupling $\ket{1} (\equiv \ket{\uparrow})$ to \ket{X} thus allows for exciton-mediated spin manipulation.
\

In a frame rotating with the laser frequency $\omega_l$, and after making a rotating wave approximation (RWA), we obtain the Hamiltonian of such a QD in the basis $\{ \ket{0}, \ket{1}, \ket{X} \}$ ($\hbar = 1$)
\begin{equation}
H_S = \Delta \kb{X}{X} + \frac{\Omega}{2} \left( \kb{1}{X} +\mathrm{h.c.} \right),
\label{eqn:2ls_hamiltonian}
\end{equation}
where  h.c. denotes the Hermitian conjugate, $\Delta \equiv  \omega_0 - \omega_l$ is the detuning between $\omega_l$ and the exciton creation energy $\omega_0$, and  $\Omega$ is the laser-QD coupling strength.

Hamiltonian (\ref{eqn:2ls_hamiltonian}) provides a valid description of the driven QD, regardless of whether a classical laser field or a quantum mechanical laser mode in a coherent state is assumed.~\cite{cohen-tannoudji92} However, the eigenstates of the joint system in a fully quantum mechanical treatment contain the photon number $N$ of the laser mode\footnote{For clarity, we use the simpler number state rather than a coherent state, which suffices for this purpose~.\cite{cohen-tannoudji92}}, giving rise to the following representation in the so-called `dressed basis':~\cite{cohen-tannoudji92}
\begin{eqnarray}
\ket{-}_N  &= &\cos \theta \ket{1, N+1} - \sin\theta \ket{X, N}, \label{eqn:dressed_minus} \\
\ket{+}_N & =  &\sin \theta \ket{1, N+1} + \cos \theta \ket{X, N}, \label{eqn:dressed_plus}\\
\theta &  = & \frac{1}{2} \arctan\left( \frac{\Omega}{\Delta}\right). \label{eqn:dressed_theta}
\end{eqnarray}
The states $\ket{-}_N$ and $\ket{+}_N$ are conveniently grouped and referred to as a manifold $\mathcal{M}(N)$. In this picture, there is a ladder of manifolds, each manifold separated from its neighbours by the energy of a laser photon $\omega_l$.~\cite{cohen-tannoudji92} Henceforth, we shall label the $\ket{-}_N$ state simply as  \ket{-} whenever information about the photon number in the laser mode is not needed.

\begin{figure}
\begin{center}
\includegraphics[width=0.4\textwidth]{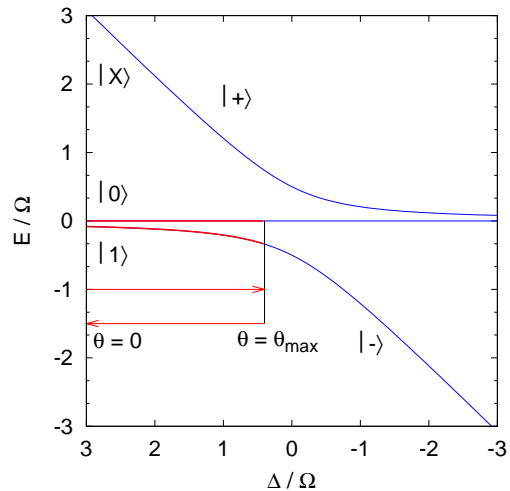}
\caption{(Color online) \textbf{Energy of the dressed states:} shown as a function of the detuning $\Delta$ in arbitrary units. The \ket{-} state tends to \ket{1} and \ket{+} to \ket{X} for a large positive detuning. At resonance, $\Delta = 0$, the dressed states are an equal superposition of the bare basis states. The decoupled \ket{0} state is also shown, its energy remaining constant. \textbf{Adiabatic following:} population in the \ket{0} and \ket{1} states follows the respective instantaneous eigenstates if the mixing angle $\theta$ changes sufficiently slowly.  \ket{0} is unaffected by the laser pulse but \ket{1} goes into \ket{-} and back. Due to the energy shift of the \ket{-} state, a dynamical phase accumulates relative to \ket{0}.}
\label{fig:adiabatic_following}
\end{center}
\end{figure}

Based on Hamiltonian (\ref{eqn:2ls_hamiltonian}), a $R_Z(\phi)$ gate ($\ket{0} \to \ket{0}$, $\ket{1} \to e^{i \phi}\ket{1}$) can be performed in two different ways. First, a resonant Rabi flop allows one to achieve any angle of rotation provided control of the driving laser phase is possible.~\cite{lovett06}
Second, slow switching of an off-resonance laser beam can be used to achieve adiabatic following of instantaneous system eigenstates. As depicted in Fig \ref{fig:adiabatic_following}, population initally in \ket{1} follows the \ket{-} state as $\theta$ goes from zero to some value $\theta_{max}$ and returns back to \ket{1} as $\theta$ goes back to zero. During this process, the energy difference between \ket{-} and \ket{0} ,
\begin{equation}
E_m = \frac{1}{2} \left( \Delta - \sqrt{\Delta^2 + \Omega^2} \right),
\end{equation}
causes an accummulation of phase of \ket{1} relative to \ket{0}.  This kind of adiabatic following can be achieved by using a laser pulse with slowly changing Gaussian field amplitude $\Omega(t) = \Omega_0 \exp[-(t/\tau)^2]$ with a constant detuning $\Delta$.

\section{Radiative decay}

Excitonic lifetimes up to a nanosecond have been reported.~\cite{borri01} Therefore, the problem of spontaneous photon emission has previously often been assumed to be insignificant compared with other decoherence channels.~\cite{calarco03,lovett05}
However, to achieve adiabaticity, our quantum gate must be performed slowly and the finite excitonic lifetime becomes a relevant factor in limiting performance.
Spontaneous emission then causes qubit dephasing, although it does not cause qubit relaxation since \ket{0} is always uncoupled to the photon emission process.

Dynamic gates also suffer from such decoherence since any population in the excited state is susceptible to radiative decay at the rate of the inverse natural lifetime $\Gamma_0$. Let $\Xi$ denote the overall `number of expected decays' for a square $2 \pi$ pulse that is defined by the integral of the excitonic population over the pulse duration.  For undamped, resonant Rabi oscillations, the population of the excited state obeys $\langle \kb{X}{X} \rangle = \sin^2(\Omega / 2 t)$, so that
\begin{equation}
\Xi = \Gamma_0 \int_0^{2 \pi / \Omega} \sin^2(\Omega t / 2) \mathrm{d}t = \frac{\pi}{\Omega} \Gamma_0.
\label{eqn:xi_omega}
\end{equation}
In principle a stronger driving reduces $\Xi$. 
However, increasing the pulse amplitude is technically demanding and can lead to population leakage from the trion subspace.

\begin{figure*}
\begin{center}
\includegraphics[width=0.65\textwidth]{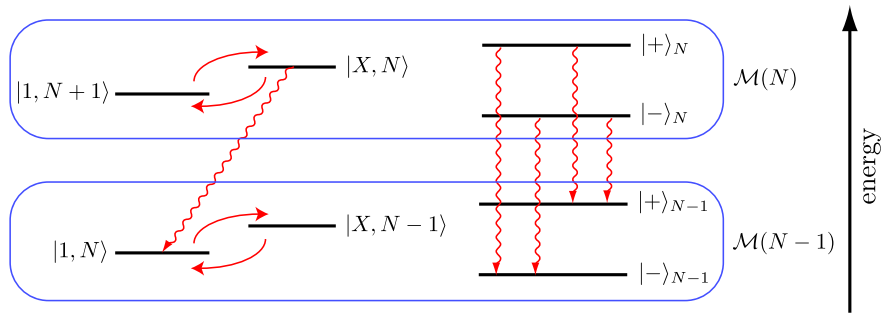}
\caption{Radiative decay from manifold $\mathcal{M}(N)$ to $\mathcal{M}(N-1) $. \textbf{Left:} the uncoupled basis. The energy between the two states in each manifold is the detuning $\Delta$ and the solid arrows correspond to absorption and stimulated emission processes, whereas the wavy arrows denote spontaneous emission. \textbf{Right:} Allowed spontaneous emission transitions between the dressed states. The energetic splitting in each manifold is the effective Rabi frequency $\Omega' = \sqrt{\Delta^2 + \Omega^2}$ and the spacing between adjacent manifolds is the laser frequency $\omega_l$. The left emission process is at frequency $\omega_l + \Omega'$, the two centre-lines emit at the frequency of the laser and the right at $\omega_l - \Omega'$; this is the Mollow triplet well known in quantum optics.~\cite{mollow69}}
\label{fig:dressed_state_transitions}
\end{center}
\end{figure*}

\begin{figure}
\begin{center}
\includegraphics[width=0.5\textwidth]{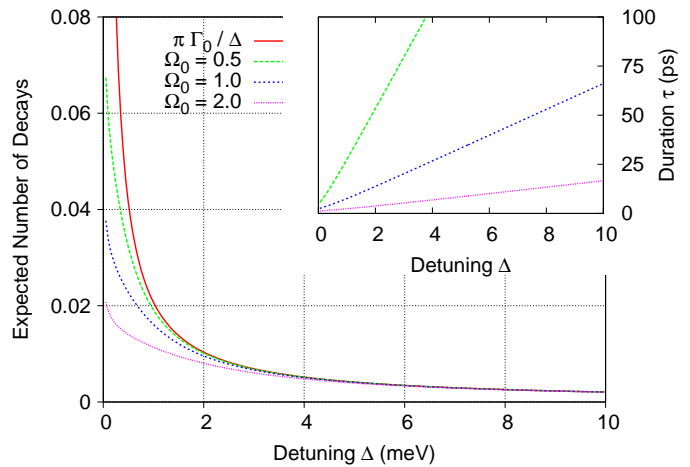}
\caption{(Color online) Any significant population of the excitonic state makes decay events inevitable. The expected number of decays (for $\Gamma_0 = 0.01~\text{ps}^{-1}$) during an adiabatic $R_Z(\pi)$ gate is shown as a function of $\Delta$.  We depict the analytical prediction (red) along with full numerical solutions,  for which $\Omega_0$ is given in meV. The inset shows the corresponding pulse durations $\tau$ in ps. }
\label{fig:number_decays}
\end{center}
\end{figure}

In the adiabatic scheme, radiative transition rates between dressed states [Eqs. (\ref{eqn:dressed_minus}) and (\ref{eqn:dressed_plus})] are required to describe spontaneous photon emission processes. These are obtained analogously to the case of an atom interacting with a laser pulse.~\cite{cohen-tannoudji92} As can be seen in Fig \ref{fig:dressed_state_transitions}, the total decay rate $\Gamma_m$ from \ket{-} into the adjacent manifold is given by the sum of two processes $\ket{-}_N \to \ket{-}_{N-1}$ and $\ket{-}_N \to \ket{+}_{N-1}$,~\cite{cohen-tannoudji92}
\begin{equation}
\Gamma_m = \Gamma_0 \sin^2 \theta \cos^2 \theta + \Gamma_0 \sin^4 \theta  = \Gamma_0  \sin^2 \theta.
\label{eqn:minus_decay}
\end{equation}
Obviously, $\Gamma_m$ decreases as $\theta$ gets smaller. On the other hand, a smaller $\theta$ entails a prolonged gating time. By using Eq. (\ref{eqn:dressed_theta}) and performing a series expansion in $\Omega / \Delta$ around $\Omega / \Delta = 0$, we obtain
\begin{equation}
\Gamma_m =  \frac{\Gamma_0}{4} \left( \frac{\Omega}{\Delta} \right)^2 + \mathcal{O} \left(\frac{\Omega}{\Delta}  \right)^4.
\end{equation}
A similar expansion for the energy shift $E_m$ of the \ket{-} state yields
\begin{equation}
E_m = \frac{\Omega}{4} \left( \frac{\Omega}{\Delta} \right) +  \mathcal{O} \left( \frac{\Omega}{\Delta} \right)^3.
\end{equation}
The time required for a $R_Z(\pi)$ operation is $\tau = \pi / E_m$. The expected number of decays during the gate is simply the product of time and decay rate, yielding
\begin{equation}
\Xi = \frac{\pi}{\Delta} \Gamma_0.
\label{eqn:xi_delta}
\end{equation}
Hence $\Xi$ decreases with increasing $\Delta$ at the cost a longer gating duration $\tau$. We verify this result in Fig. \ref{fig:number_decays}, which compares the result Eq. (\ref{eqn:xi_delta}) with a full numerical simulation, showing an excellent agreement whenever $\Delta \gtrsim 2 \Omega$. We have checked our results with an appropriate Markovian ME equation \footnote{$\dot{\rho} = - i [H_S, \rho] + \Gamma_0 \left(\sm \rho \sp - \frac{1}{2} (\sp \sm \rho + \rho \sp \sm)  \right)$}, which reconfirms that further detuning the laser reduces radiative decay in the adiabatic scheme, while it takes a stronger coupling $\Omega$ to achieve the same for the dynamic gate.

\section{Phonon interactions}

Interactions with vibrational modes of the surrounding lattice cause twofold decoherence. First, they dephase the qubit in a similar way to radiative decay and second they can lead to gate failure through state relaxation. We focus on deformation potential coupling to acoustic phonons as the main phonon induced dephasing mechanism for slow excitonic processes,~\cite{mahan00, krummheuer02} since it dominates over the much weaker piezoelectric coupling.~\cite{pazy02, krummheuer02}

In addition to the electron spin qubit, the trion state \ket{X} also consists of a strongly bound exciton. Therefore, its interaction with phonons is given by~\cite{mahan00, krummheuer02}
\begin{equation}
H_{ep} =  \kb{X}{X} \sum_\vec{q} g_\vec{q} \left(a_\vec{q}  +  \text{h.c.} \right),
\label{eqn:ph_ham_int}
\end{equation}
where $a_\vec{q}$ is the annihilation operator for a phonon with wavevector $\vec{q}$ and $g_\vec{q}$ is an effective excitonic coupling strength $g_\vec{q} \equiv (M^e_\vec{q} \mathcal{P}[\psi^{e}(\vec{r})] - M^h_\vec{q} \mathcal{P}[\psi^{h}(\vec{r})])$.~\cite{krummheuer02} The $M^{e/h}_\vec{q} = D_{e/h} \abs{\vec{q}} \sqrt{\hbar / (2 \mu V  \omega_\vec{q}})$ are the deformation potential coupling strengths for electrons and holes,~\cite{mahan00} with $\mu$ being the mass density, $V$ the lattice volume, and $D_{e/h}$ the respective electron and hole coupling constants. $\mathcal{P}[\psi^{e/h}]$ denotes the form factor of the electron / hole wavefunction.

We proceed by transforming Eq. (\ref{eqn:ph_ham_int}) into the diagonal basis of $H_S$ (Eq. (\ref{eqn:2ls_hamiltonian})) and by then writing it in the interaction picture with respect to both $H_S$ and $H_B = \sum_\vec{q} \omega_\vec{q}  \cre{a}_\vec{q}  \an{a}_\vec{q}$, yielding,
\begin{eqnarray}
\tilde{H}_I(t) =  \sum\limits_{\omega',\vec{q}} \left( P_{\omega'} e^{-i \omega' t}  + \mathrm{h.c}  \right)  g_\vec{q} \left(\an{a}_\vec{q} e^{-i \omega_\vec{q} t} + \mathrm{h.c.} \right),
\label{eqn:ph_ham_diag}
\end{eqnarray}
where $\omega' \in \{ 0, \Lambda \}$,  $P_0 = \cos^2 \theta \kb{+}{+} + \sin^2 \theta \kb{-}{-}$ and $P_\Lambda = -\sin \theta \cos \theta \kb{-}{+}$. $\Lambda = \sqrt{\Omega^2 + \Delta^2}$ is the energy difference between the dressed states.

The ME is derived in the usual way~\cite{breuer02} by integrating the von Neumann equation for the density matrix $\varrho$ of the joint system and tracing over the phonon modes. This results in an integro-differential equation for the qubit density matrix $\rho$:
\begin{equation}
\dot{\rho} = - \int\limits_0^t \mathrm{d}t' \mathrm{tr}_{ph} \left( [\tilde{H}_I(t), [\tilde{H}_I(t'), \varrho(t')]] \right).
\end{equation}
The Born-Markov approximation is now performed, which relies on two assumptions. First, there is no backaction from the small system on the much larger bath, meaning the joint density matrix factorizes at all times $\varrho = \rho \otimes \rho_B$. Second, the bath relaxation is assumed to be rapid and so we replace $\rho(t')$ by $\rho(t)$, thereby neglecting any memory effects.~\cite{breuer02} Further, we assume that system dynamics occurs on a timescale much faster than the decoherence processes, $J(\Lambda) \ll \Lambda$, allowing us to perform a RWA and leading to a ME in Lindblad form:~\cite{breuer02, stace05}
\begin{equation}
\dot{\rho} = J(\Lambda)\left( [N(\Lambda) + 1] D[P_{\Lambda}]\rho +  N(\Lambda) D[P_{\Lambda}^{\dagger}]\rho \right).
\label{eqn:ph_me}
\end{equation}
$D[L]\rho \equiv  L \rho L^{\dagger} - 1/2 (L^{\dagger} L \rho + \rho L^{\dagger} L)$ is the `dissipator' of the ME, \mbox{$N(\omega) = \left(\exp (\omega / k_B T) - 1 \right)^{-1}$} describes the thermal occupation of the phonon modes, and $J(\omega)$ is the phonon spectral density:
\begin{equation}
J(\omega) = 2 \pi \sum_{\vec{q}} \abs{g_\vec{q}}^2 \delta (\omega - \omega_\vec{q}).
\end{equation}

The Lindblad operator $P_0$ has been dropped because the spectral density vanishes for $\omega = 0$. Consequently, the ME in the Born-Markov approximation provides a suitable description of phonon-assisted transitions but does not describe pure dephasing~\cite{krummheuer02, pazy02} in the diagonal basis. Pure dephasing is an intrinisically non-Markovian process~\cite{roszak05} that can be interpreted as a partial `which-way-measurement' of the qubit by the environment, thus causing decoherence.~\cite{roszak06} However, it can be adiabatically eliminated if system dynamics proceed no faster than at a characteristic timescale of the order of $1~\text{ps}$.~\cite{alicki04, calarco03, roszak05}

\begin{figure}
\begin{center}
\includegraphics[width=0.5\textwidth]{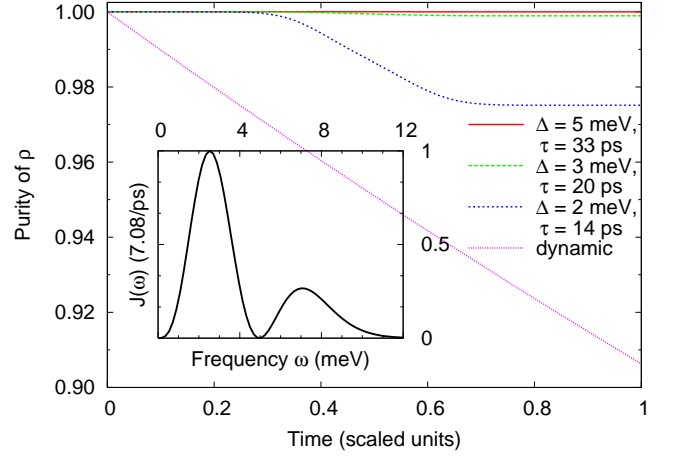}
\caption{(Color online) Phonons affect dynamic and adiabatic quantum operations differently. To characterize decoherence, we present the purity of the system's density matrix during a $R_z(\pi)$  operation for deformation potential coupling at $T=5 ~\text{K}$. The dynamic gate uses $\Omega=0.1~\text{meV}$ and time is shown in units of $2 \pi / \Omega$. Adiabatic gates are performed with $\Omega_0=1~\text{meV}$ and the units of $\Delta$ are meV.
We have scaled the time axis by dividing by $6 \tau$ and adding $0.5$ (thus mapping the actual integration interval of $- 3 \tau$ to $3 \tau$ to the scale of this figure).
The inset shows the deformation potential spectral density of GaAs [Eq. (\ref{eqn:spectral_density})]. }
\label{fig:phonons}
\end{center}
\end{figure}

The calculation of the spectral density $J (\omega)$ requires a microscopic model for the carrier wavefunctions. The precise wavefunction shape is not important \footnote{The dynamic gate uses small $\Lambda$, where $\mathcal{P}[\psi^{e/h}(\vec{r})] \approx 1$ while the adiabatic gate operates best beyond the spectral cut-off of phonon modes.} so we proceed by using ground state solutions to a harmonic confinement potential of strength $162~\text{meV}$ at $2.5~\text{nm}$ from the centre of the dot:~\cite{nazir05} $\psi_{e/h}(\vec{r}) = (d_{e/h} \sqrt{\pi})^{-\frac{3}{2}} \exp(- r^2 / (2 d^2_{e/h}))$, for which  $d_{e/h} = (\hbar / \sqrt{m_{e/h} c})^{1/2}$ and $c=8.3\times10^{-3}~\text{J/m}^2$. For GaAs:~\cite{krummheuer02, pazy02} $D_e=14.6~\text{eV}$, $D_h=4.8~\text{eV}$, $\mu=5.3~\text{g/cm}^3$ and $c_s=4.8 \times10^5~\text{cm/s}$. The effective electron and hole masses are $m_e=0.067~m_0$,  $m_h=0.34~m_0$. Assuming a linear phonon dispersion $\omega_\vec{q}=c_s \abs{\vec{q}}$, the spectral density then takes a super-Ohmic form (shown in the inset of  Fig. \ref{fig:phonons})
\begin{equation}
 J(\omega)  =  \frac{D_e^2 \omega^3}{2 \mu c_s^5} \left( e^{-\frac{\omega^2}{\omega_{e}^2}}  -  2 \frac{D_h}{D_e} e^{-\frac{\omega^2}{\omega_{eh}^2}}  + \frac{D_h^2}{D_e^2}  e^{-\frac{\omega^2}{\omega_{h}^2}} \right).
\label{eqn:spectral_density}
\end{equation}
The exponential cut-off terms on the RHS of Eq. (\ref{eqn:spectral_density}) are related to the finite size of the QD and filter out high frequency phonons with wavelengths too short to interact with the dot. The cut-off frequencies are: $\omega_{e,h} = \sqrt{2} c_s / d_{e,h}$ and $\omega_{eh} = 2 c_s / \sqrt{d_e^2 + d_h^2} $.

By examining the structure of Eqs. (\ref{eqn:ph_me}) and  (\ref{eqn:spectral_density}) we conclude that the influence of phonon-induced decoherence becomes small as $\Lambda$  approaches zero and can be exponentially suppressed for values of $\Lambda$ beyond the spectral cut-off $\omega_{e/h}$. Outside the ultrafast regime only the first route of decreasing $\Lambda$ via $\Omega$ lends itself to the dynamic scheme but this is contrary to the requirement of avoiding radiative decay. On the other hand, operation beyond the cut-off is realizable for an adiabatic gate with a large enough detuning. Using the purity of the system's density matrix, $\mathrm{tr} [\rho^2]$, to characterize system decoherence, Fig. \ref{fig:phonons} shows that 
phonon decoherence can indeed be overcome with the adiabatic approach and that in general performance improves as the detuning increases \footnote{While not shown here, the much weaker piezoelectric coupling~\cite{mahan00} is also suppressed since its Ohmic spectral density is subject to the same cut-off frequencies: $J(\omega)  =  \frac{P^2 \omega^3}{2 \mu c_s^3} \left( e^{-\omega^2/\omega_{e}^2}  -  2 e^{-\omega^2/\omega_{eh}^2}  + e^{-\omega^2/\omega_{h}^2} \right)$ (for LA phonons, for TA phonons it is even lower).}.

\section{Landau-Zener transitions}

Landau-Zener (LZ) transitions are non-adiabatic transitions between eigenstates approaching an anticrossing;~\cite{wubs05} they are a source of error in the adiabatic scheme since they cause population to transfer to excited states, subjecting it to fast decoherence during and after the gate. LZ transitions can be suppressed by using longer gating times. Contrary to the adiabaticity conditions derived in Refs. \onlinecite{calarco03} and \onlinecite{lovett05}, which are only valid for a linear sweep through resonance, we take a more general approach.

To derive an adiabaticity condition consider the 2LS Hamiltonian (\ref{eqn:2ls_hamiltonian}). The transformation to the basis of instantaneous eigenstates,
\begin{equation}
U(\theta) =
\left(\begin{array}{cc}
\cos \theta & - \sin \theta \\
\sin \theta & \cos \theta
\end{array}\right),
\end{equation}
is time-dependent. The Hamiltonian transforms as $\tilde{H} = U^{\dagger} H U + i \left(  \ddt  U^{\dagger} \right) U$, yielding
\begin{equation}
\tilde{H} = \lambda^{-} \kb{-}{-} + \lambda^{+} \kb{+}{+} + \dot{\theta} \left(  i \kb{-}{+} + \mathrm{h.c.}  \right),
\end{equation}
where $\lambda^{\pm} = 1/2 (\Delta \pm \sqrt{\Delta^2 + \Omega^2})$ denote the instantaneous eigenenergies of the states \ket{-} and \ket{+}. The off-diagonal terms give rise to LZ transitions, which are therefore suppressed if
\begin{equation}
\dot{\abs{\theta}} \ll \abs{\lambda^{+} - \lambda^{-}}.
\end{equation}
Using Eq. (\ref{eqn:dressed_theta}) for $\theta$, we obtain a general condition for adiabaticity:
\begin{equation}
\frac{\dot{\Omega}\Delta - \Omega \dot{\Delta}} {2 (\Delta^2 + \Omega^2)^{3/2}} \ll 1.
\label{eqn:adiabaticity_condition}
\end{equation}
For $\Omega = \Omega_0 \exp[-(t/\tau)^2]$ and constant $\Delta$, we deduce that inequality (\ref{eqn:adiabaticity_condition}) can be satisfied by demanding
\begin{equation}
\Omega_0 / \Delta^2 \ll \tau.
\end{equation}
Adiabatic following is therefore always achieved in the limit $\Omega \ll \Delta$, where $\tau \propto \Delta$ over a broad range of parameter space.

\section{Overall Gate Fidelity}

In order to bring together the results so far, we now calculate the fidelity of the $R_Z(\pi)$ gate with a combined ME describing both spontaneous photon emission and phonon-induced processes. The gate is performed on the input state $\ket{\psi_+} = (\ket{0} + \ket{1}) / \sqrt{2}$ and ideally produces $\ket{\psi_-} = (\ket{0} - \ket{1}) / \sqrt{2}$ as its output state. The fidelity is defined as $\mathcal{F} = \bra{\psi_-} \rho \ket{\psi_-}$, where $\rho$ is the density matrix of the system after the gate has finished. Again, we use the material parameters of GaAs and assume a radiative decay rate of $\Gamma_0 = 0.01~\text{ps}^{-1}$.

\begin{figure}
\begin{center}
\includegraphics[width=0.5\textwidth]{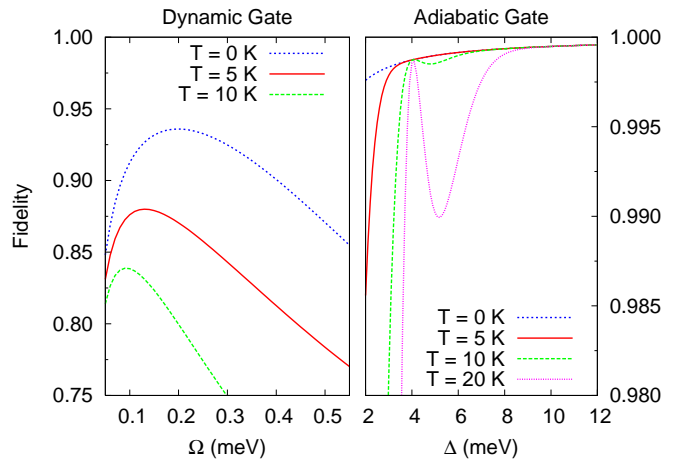}
\caption{(Color online) Comparison of the dynamic versus the adiabatic gate for a $R_Z(\pi)$ operation by showing the overall gate fidelity. \textbf{Left:} fidelity of the dynamic operation as a function of the coupling strength $\Omega$. \textbf{Right:} fidelity of the adiabatic operation for a fixed $\Omega_0 = 1~\text{meV}$ as a function of the detuning. }
\label{fig:1qd_fidelity}
\end{center}
\end{figure}

The left panel of Fig. \ref{fig:1qd_fidelity} shows the fidelity of the dynamic gate as a function of $\Omega$ for different temperatures. Towards small values of $\Omega$, the fidelity is limited by the finite excitonic lifetime, whereas phonon-induced processes become important for larger values of $\Omega$. The highest fidelity here is below $0.95$ at absolute zero and decreases at finite temperatures.
For much higher values of $\Omega$ beyond the phonon spectral cutoff ($\sim 10$~meV), we would expect the fidelity to improve rapidly, neglecting the effects of pure dephasing. However, as outlined above, our approach is not suited to studying the regime of ultrafast driving, and we refrain from showing actual values at large $\Omega$.

The fidelity of the adiabatic gate as a function of $\Delta$ for a fixed value of $\Omega_0 = 1~\text{meV}$ is shown in the right panel of Fig. \ref{fig:1qd_fidelity}. Towards larger values of $\Delta$, $\mathcal{F}$ increases and asymptotically approaches unity as radiative decay is more effectively suppressed. Phonon-induced processes are temperature dependent and predominantly occur for small values of $\Delta$. The intermediate peak visible at higher temperatures is related to the dip in the spectral density of the deformation potential (see inset of Fig. \ref{fig:phonons}).

\section{Conclusions}

We have performed a comprehensive Markovian decoherence study of an exciton-mediated spin phase gate in quantum dots. 
Within an adiabatic scheme all principal sources of decoherence can be effectively suppressed in the same limit of weak off-resonant laser driving, and a phase gate fidelity of 0.999 or better may be possible even at finite temperature.
In contrast, for similar driving amplitudes, a dynamic gate with typical operating parameters suffers strong decoherence that leads to somewhat lower gate fidelity.

\begin{acknowledgements}
We thank Avinash Kolli, Tom Stace and Jay Gambetta for valuable and engaging discussions. EMG acknowledges support from the Marie Curie Early Stage Training network ÔQIPESTÕ (MEST-CT-2005-020505) and the QIPIRC (No. GR/S82176/01) for support. AN is supported by Griffith University, the State of Queensland, and the Australian Research Council Special Research Centre for Quantum Computer Technology. BWL and SCB acknowledge support from the Royal Society.
\end{acknowledgements}

\end{document}